\documentstyle[12pt]{article}
\newcommand{\nc}{\newcommand}
\nc{\renc}{\renewcommand}

\nc{\etal}{\mbox{\it et al. }}
\nc{\ie}{{\it i.e.}}
\nc{\eg}{{\it e.g.}}

\renc{\thefootnote}{\arabic{footnote}}
\nc{\capt}[1]{{\bf Figure.} {\small\sl #1}}

\nc{\eqs}[2]{\mbox{Eqs.~(\ref{#1},\,\ref{#2})}}
\nc{\eq}[1]{\mbox{Eq.~(\ref{#1})}}

\nc{\figs}[2]{\mbox{Figs.~(\ref{#1},\,\ref{#2})}}
\nc{\fig}[1]{\mbox{Fig~.(\ref{#1})}}

\nc{\tag}[1]{\label{#1} \marginpar{{\footnotesize #1}}}
\nc{\mtag}[1]{\label{#1} \mbox{\marginpar{{\footnotesize #1}}}}
\renc{\baselinestretch}{1.2}
\newlength{\overeqskip}
\newlength{\undereqskip}
\nc{\be}[1]{\begin{equation} \mbox{$\label{#1}$}}
\nc{\bea}[1]{\begin{eqnarray} \mbox{$\label{#1}$}}
\nc{\Section}[2]{\section{#2}\label{#1}}
\nc{\Bibitem}[1]{\bibitem{#1}}
\nc{\Label}[1]{\label{#1}}

\nc{\eea}{\vspace{\undereqskip}\end{eqnarray}}
\nc{\ee}{\vspace{\undereqskip}\end{equation}}
\nc{\bdm}{\begin{displaymath}}
\nc{\edm}{\end{displaymath}}
\nc{\dpsty}{\displaystyle}
\nc{\bc}{\begin{center}}
\nc{\ec}{\end{center}}
\nc{\ba}{\begin{array}}
\nc{\ea}{\end{array}}
\nc{\bab}{\begin{abstract}}
\nc{\eab}{\end{abstract}}
\nc{\btab}{\begin{tabular}}
\nc{\etab}{\end{tabular}}
\nc{\bit}{\begin{itemize}}
\nc{\eit}{\end{itemize}}
\nc{\ben}{\begin{enumerate}}
\nc{\een}{\end{enumerate}}
\nc{\bfig}{\begin{figure}}
\nc{\efig}{\end{figure}}
\nc{\arreq}{&\!=\!&}
\nc{\arrmi}{&\!-\!&}
\nc{\arrpl}{&\!+\!&}
\nc{\arrap}{&\!\!\!\approx\!\!\!&}
\nc{\non}{\nonumber\\*}
\nc{\align}{\!\!\!\!\!\!\!\!&&}

\def\lsim{\; \raise0.3ex\hbox{$<$\kern-0.75em
      \raise-1.1ex\hbox{$\sim$}}\; }
\def\gsim{\; \raise0.3ex\hbox{$>$\kern-0.75em
      \raise-1.1ex\hbox{$\sim$}}\; }
\nc{\DOT}{\hspace{-0.08in}{\bf .}\hspace{0.1in}}
\nc{\Laada}{\hbox {$\sqcap$ \kern -1em $\sqcup$}}
\nc\loota{{\scriptstyle\sqcap\kern-0.55em\hbox{$\scriptstyle\sqcup$}}}
\nc\Loota{{\sqcap\kern-0.65em\hbox{$\sqcup$}}}
\nc\laada{\Loota}
\nc{\qed}{\hskip 3em \hbox{\BOX} \vskip 2ex}

\nc{\real}{{\rm I \! R}}
\nc{\Z}{{\sf Z \!\!\! Z}}
\nc{\complex}{{\rm C\!\!\! {\sf I}\,\,}}
\def\bigid{\leavevmode\hbox{\small1\kern-3.8pt\normalsize1}}
\def\id{\leavevmode\hbox{\small1\kern-3.3pt\normalsize1}}
\nc{\slask}{\!\!\!/}
\nc{\bis}{{\prime\prime}}
\nc{\pa}{\partial}
\nc{\na}{\nabla}
\nc{\ra}{\rangle}
\nc{\la}{\langle}
\nc{\goto}{\rightarrow}
\nc{\swap}{\leftrightarrow}

\nc{\EE}[1]{ \mbox{$\cdot10^{#1}$} }
\nc{\abs}[1]{\left|#1\right|}
\nc{\at}[2]{\left.#1\right|_{#2}}
\nc{\norm}[1]{\|#1\|}
\nc{\abscut}[2]{\Abs{#1}_{\scriptscriptstyle#2}}
\nc{\vek}[1]{{\rm\bf #1}}
\nc{\integral}[2]{\int\limits_{#1}^{#2}}
\nc{\inv}[1]{\frac{1}{#1}}
\nc{\dd}[2]{{{\partial #1}\over{\partial #2}}}
\nc{\ddd}[2]{{{{\partial}^2 #1}\over{\partial {#2}^2}}}
\nc{\dddd}[3]{{{{\partial}^2 #1}\over
	{\partial #2 \partial #3}}}
\nc{\dder}[2]{{{d #1}\over{d #2}}}
\nc{\ddder}[2]{{{d^2 #1}\over{d {#2}^2}}}
\nc{\dddder}[3]{{d^2 #1}\over
	{d #2 d #3}}
\nc{\dx}[1]{d\,^{#1}x}
\nc{\dy}[1]{d\,^{#1}y}
\nc{\dz}[1]{d\,^{#1}z}
\nc{\dl}[1]{\frac{d\,^{#1}l}{(2\pi)^{#1}}}
\nc{\dk}[1]{\frac{d\,^{#1}k}{(2\pi)^{#1}}}
\nc{\dq}[1]{\frac{d\,^{#1}q}{(2\pi)^{#1}}}

\nc{\cc}{\mbox{$c.c.$ }}
\nc{\hc}{\mbox{$h.c.$ }}
\nc{\cf}{cf.\ }
\nc{\erfc}{{\rm erfc}}
\nc{\Tr}{{\rm Tr\,}}
\nc{\tr}{{\rm tr\,}}
\nc{\pol}{{\rm pol}}
\nc{\sign}{{\rm sign}}
\nc{\bfT}{{\bf T }}

\nc{\cA}{{\cal A}}
\nc{\cB}{{\cal B}}
\nc{\cD}{{\cal D}}
\nc{\cE}{{\cal E}}
\nc{\cG}{{\cal G}}
\nc{\cH}{{\cal H}}
\nc{\cL}{{\cal L}}
\nc{\cO}{{\cal O}}
\nc{\cT}{{\cal T}}
\nc{\cN}{{\cal N}}
\nc{\rvac}[1]{|{\cal O}#1\rangle}
\nc{\lvac}[1]{\langle{\cal O}#1|}
\nc{\rvacb}[1]{|{\cal O}_\beta #1\rangle}
\nc{\lvacb}[1]{\langle{\cal O}_\beta #1 |}
\nc{\bb}{\bar{\beta}}
\nc{\bt}{\tilde{\beta}}
\nc{\ctH}{\tilde{\cal H}}
\nc{\chH}{\hat{\cal H}}
\nc{\1}{\aa}
\nc{\2}{\"{a}}
\nc{\3}{\"{o}}
\nc{\4}{\AA}
\nc{\5}{\"{A}}
\nc{\6}{\"{O}}
\nc{\al}{\alpha}
\nc{\g}{\gamma}
\nc{\Del}{\Delta}
\nc{\e}{\epsilon}
\nc{\eps}{\epsilon}
\nc{\lam}{\lambda}
\nc{\om}{\omega}
\nc{\Om}{\Omega}
\nc{\ve}{\varepsilon}
\nc{\mn}{{\mu\nu}}
\nc{\k}{\kappa}
\nc{\vp}{\varphi}

\nc{\advp}[3]{{\it  Adv.\ in\ Phys.\ }{{\bf #1} {(#2)} {#3}}}
\nc{\annp}[3]{{\it  Ann.\ Phys.\ (N.Y.)\ }{{\bf #1} {(#2)} {#3}}}
\nc{\apl}[3]{{\it  Appl. Phys. Lett. }{{\bf #1} {(#2)} {#3}}}
\nc{\apj}[3]{{\it  Ap.\ J.\ }{{\bf #1} {(#2)} {#3}}}
\nc{\apjl}[3]{{\it  Ap.\ J.\ Lett.\ }{{\bf #1} {(#2)} {#3}}}
\nc{\app}[3]{{\it Astropart.\ Phys.\ }{{\bf #1} {(#2)} {#3}}}
\nc{\cmp}[3]{{\it  Comm.\ Math.\ Phys.\ }{{ \bf #1} {(#2)} {#3}}}
\nc{\cqg}[3]{{\it  Class.\ Quant.\ Grav.\ }{{\bf #1} {(#2)} {#3}}}
\nc{\epl}[3]{{\it  Europhys.\ Lett.\ }{{\bf #1} {(#2)} {#3}}}
\nc{\ijmp}[3]{{\it Int.\ J.\ Mod.\ Phys.\ }{{\bf #1} {(#2)} {#3}}}
\nc{\ijtp}[3]{{\it Int.\ J.\ Theor.\ Phys.\ }{{\bf #1} {(#2)} {#3}}}
\nc{\jmp}[3]{{\it  J.\ Math.\ Phys.\ }{{ \bf #1} {(#2)} {#3}}}
\nc{\jpa}[3]{{\it  J.\ Phys.\ A\ }{{\bf #1} {(#2)} {#3}}}
\nc{\jpc}[3]{{\it  J.\ Phys.\ C\ }{{\bf #1} {(#2)} {#3}}}
\nc{\jap}[3]{{\it J.\ Appl.\ Phys.\ }{{\bf #1} {(#2)} {#3}}}
\nc{\jpsj}[3]{{\it J.\ Phys.\ Soc.\ Japan\ }{{\bf #1} {(#2)} {#3}}}
\nc{\lmp}[3]{{\it Lett.\ Math.\ Phys.\ }{{\bf #1} {(#2)} {#3}}}
\nc{\mpl}[3]{{\it  Mod.\ Phys.\ Lett.\ }{{\bf #1} {(#2)} {#3}}}
\nc{\ncim}[3]{{\it  Nuov.\ Cim.\ }{{\bf #1} {(#2)} {#3}}}
\nc{\np}[3]{{\it  Nucl.\ Phys.\ }{{\bf #1} {(#2)} {#3}}}
\nc{\pr}[3]{{\it Phys.\ Rev.\ }{{\bf #1} {(#2)} {#3}}}
\nc{\pra}[3]{{\it  Phys.\ Rev.\ A\ }{{\bf #1} {(#2)} {#3}}}
\nc{\prb}[3]{{\it  Phys.\ Rev.\ B\ }{{{\bf #1} {(#2)} {#3}}}}
\nc{\prc}[3]{{\it  Phys.\ Rev.\ C\ }{{\bf #1} {(#2)} {#3}}}
\nc{\prd}[3]{{\it  Phys.\ Rev.\ D\ }{{\bf #1} {(#2)} {#3}}}
\nc{\prl}[3]{{\it Phys\ Rev.\ Lett.\ }{{\bf #1} {(#2)} {#3}}}
\nc{\pl}[3]{{\it  Phys.\ Lett.\ }{{\bf #1} {(#2)} {#3}}}
\nc{\prep}[3]{{\it Phys\. Rep.\ }{{\bf #1} {(#2)} {#3}}}
\nc{\prsl}[3]{{\it Proc.\ R.\ Soc.\ London\ }{{\bf #1} {(#2)} {#3}}}
\nc{\ptp}[3]{{\it  Prog.\ Theor.\ Phys.\ }{{\bf #1} {(#2)} {#3}}}
\nc{\ptps}[3]{{\it  Prog\ Theor.\ Phys.\ suppl.\ }{{\bf #1} {(#2)} {#3}}}
\nc{\physa}[3]{{\it  Physica\ A\ }{{\bf #1} {(#2)} {#3}}}
\nc{\physb}[3]{{\it  Physica\ B\ }{{\bf #1} {(#2)} {#3}}}
\nc{\phys}[3]{{\it Physica\ }{{\bf #1} {(#2)} {#3}}}
\nc{\rmp}[3]{{\it  Rev.\ Mod.\ Phys.\ }{{\bf #1} {(#2)} {#3}}}
\nc{\rpp}[3]{{\it Rep.\ Prog.\ Phys.\ }{{\bf #1} {(#2)} {#3}}}
\nc{\sjnp}[3]{{\it Sov.\ J.\ Nucl.\ Phys.\ }{{\bf #1} {(#2)} {#3}}}
\nc{\spjetp}[3]{{\it Sov.\ Phys.\ JETP\ }{{\bf #1} {(#2)} {#3}}}
\nc{\yf}[3]{{\it Yad.\ Fiz.\ }{{\bf #1} {(#2)} {#3}}}
\nc{\zetp}[3]{{\it Zh.\ Eksp.\ Teor.\ Fiz.\  }{{\bf #1}  {(#2)} {#3}}}
\nc{\zp}[3]{{\it Z.\ Phys.\ }{{\bf #1} {(#2)} {#3}}}
\nc{\ibid}[3]{{\sl ibid.\ }{{\bf #1} {#2} {#3}}}
\nc{\rf}[1]{(\ref{#1})}
\nc{\nn}{\nonumber \\*}
\nc{\bfB}{\bf{B}}
\nc{\bfv}{\bf{v}}
\nc{\bfx}{\bf{x}}
\nc{\bfy}{\bf{y}}
\nc{\vx}{\vec{x}}
\nc{\vy}{\vec{y}}
\nc{\oB}{\overline{B}}
\nc{\oI}{\overline{I}}
\nc{\oR}{\overline{R}}
\nc{\rar}{\rightarrow}
\nc{\ti}{\times}
\nc{\slsh}{\hskip-5pt/}
\nc{\sm}{Standard~Model~}
\nc{\MP}{M_{\rm Pl}}
\nc{\tp}{t_{\rm Pl}}
\nc{\ave}{\bar{E}}

\renc{\min}{p_{\rm min}}
\renc{\max}{p_{\rm max}}
\nc{\pmin}{p_{\rm min}}
\nc{\pmax}{p_{\rm max}}
\nc{\fo}{f_0}
\nc{\foi}{f_{0,i}\,}
\nc{\fop}{f_0^P}
\nc{\fou}{f_0^U}

\nc{\eff}{{\rm eff}}
\nc{\MT}{M_{\rm T}}
\nc{\ML}{M_{\rm L}}
\nc{\kk}{\vek{k}}
\nc{\pp}{{\rm p}}
\nc{\cb}{critical bubble~}
\nc{\cbs}{critical bubbles~}
\nc{\scb}{subcritical bubble~}
\nc{\scbs}{subcritical bubbles~}
\nc{\vv}{\\}
\begin{document}

{\title{\vskip-2truecm{\hfill {{\small 
        }}\vskip 1truecm}
{\bf Do subcritical bubbles hinder first order phase transition ?}}


{\author{
{\sc Kari Enqvist $^{1}$ }\\
{\sl\small Department of Physics, P.O. Box 9,
FIN-00014 University of Helsinki,
Finland} \\
{\sc Antonio Riotto$^{2}$ }\\
{\sl\small International School for Advanced Studies, SISSA-ISAS} \vv
{\sl\small Starda Costiera 11, I-34014, Miramare, Trieste, Italy}\vv
{\sl\small and }\vv {\sl\small Istituto Nazionale di Fisica Nucleare,}
{\sl\small Sezione di Padova, 35100 Padova, Italy} \\
{\sc Iiro Vilja$^{3}$ }\\
{\sl\small Department of Physics,
University of Turku, FIN-20500 Turku, Finland}}}

\maketitle
\vspace{2cm}
\begin{abstract}
\noindent
We consider the role played by subcritical bubbles during the
electroweak phase transition, estimate their average size, amplitude and
formation rate taking into account the crucial
role played by thermalization.
We also study the influence of subcritical bubbles on
the formation of critical bubbles in the thin wall regime and show that,
contrary to some recent claims, subcritical bubbles do not affect the
nucleation of critical bubbles in an appreciable way. From this fact we
conclude that the electroweak baryogenesis scenarios associated
with a first order electroweak phase transition still remain an attractive
possibility.
\end{abstract}
\vfil
\footnoterule
{\small $^1$enqvist@phcu.helsinki.fi};
{\small $^2$riotto@tsmi19.sissa.it. Address after November 95:
Theoretical Astrophysics Group, NASA/Fermilab, Batavia, IL60510, USA.};
{\small  $^3$vilja@utu.fi}
\thispagestyle{empty}
\newpage
\setcounter{page}{1}
\Section {intro}{Introduction}

Critical bubble nucleation during a first order electroweak phase transition
has received much attention since the discovery of the  possibility
for electroweak baryogenesis \cite{ckn}. Indeed, one of the basic
ingredients for the generation of the baryon asymmetry (apart from the
requirement of
baryon- and CP-violating interactions) is the presence of
an out-of-equilibrium state \cite{sak} which, during the
first order electroweak phase
transition with supercooling, is
attained by critical bubbles expanding in the thermal
bath of the unbroken phase.

Less attention has been paid to
the environment where the critical bubble nucleation is supposed to occur.
Since critical bubbles have a finite size, phase transitions
are highly local phenomena. Fluctuations with spatial correlations
comparable to the critical bubble size may be expected to be important
for bubble nucleation. Also, if thermal fluctuations are too large,
any perturbative scheme
could break down. In such a situation
 prediction of a first order phase transition
becomes suspect, with the possibility that the entire scenario of electroweak
baryogenesis might be  invalidated.

Although the presence of thermal fluctuations in any hot system is
undisputed, their role in the dynamics of weakly first order phase
transitions is still under debate \cite{contra}.

The idea that  statistical fluctuations around equilibrium
are spherically symmetric and have roughly a correlation volume, where
the correlation length is given by the inverse temperature dependent
mass of the Higgs field, $\xi(T)=m^{-1}(T)$,
was first discussed in \cite{gleiser}. These fluctuations are
referred to as subcritical bubbles. The
amplitude of the thermal fluctuations was estimated in ref. \cite{gleiser},
where it was concluded
that they are dominant if the Higgs mass $m_H$ is larger than $\sim 80$ GeV,
whence the  fraction which the asymmetric vacuum
occupies in the neighborhood of
the critical temperature becomes of the order of unity. Therefore it was
concluded that
critical bubbles cannot be generated due to the inhomogeneities of the
background field.
In ref. \cite{gleiser}, however, the continuous
disappearance of the subcritical
bubbles was not accounted for. This can happen in two ways: the
subcritical bubbles, being unstable configurations, tend to shrink; the
bubbles are also subject of constant thermal bombardment so that they
may disappear simply because of thermal noise. The thermalization
rate of small-amplitude configurations near the critical temperature has
been estimated in ref. \cite{ElmforsEV} for the electroweak phase
transition, and it was found that, compared with typical first order
transition times, thermalization is rather fast.

Kinetics of subcritical
bubbles has been investigated by Gelmini and Gleiser \cite{gel}, who
found, under a specific assumption about the form of the destruction
rate due to thermal noise, that for Higgs masses below $\sim$ 55 GeV
the approach to equilibrium is dominated by shrinking. Unfortunately,
for the interesting range of Higgs masses dictated by the experimental
constraints coming from LEP, $m_H> 60$ GeV \cite{lep}, their analysis is
inconclusive since the approximations adopted break down. Nucleation
induced by the growth of conglomerates of subcritical bubbles has been
discussed in \cite{EV}.

Very recently, Shiromizu {\it et al.} \cite{jap} have re-estimated the
amplitude of thermal fluctuations by calculating the typical size of
subcritical bubbles during the electroweak phase transition. They claim
that for {\it any} experimentally allowed values of the Higgs mass, the
amplitude of thermal fluctuations always exceeds the first reflection
point of the effective potential.
 From this analysis they conclude that any standard electroweak baryogenesis
scenario associated with a first order first transition cannot work.

Their starting point is the observation that at the microscopic level,
the true origin of the dominant thermal fluctuation is the perpetual
creation and annihilation of spherical subcritical bubbles. Thus one
should identify the typical size $R$ of the bubbles with the size estimated
by a statistical ensemble averaging, instead of assuming it to be the
correlation length. Since the amplitude of the thermal fluctuations
sensitively depends on the size of the subcritical bubble, a small
change in the size results in a drastic change in the details of
the phase transition. Indeed, modeling any subcritical bubbles with a
Gaussian
shape, the authors of ref.
\cite{jap} find that the
typical size $\langle R\rangle$ of a subcritical bubble
is smaller than the correlation length,
and that the amplitude of thermal fluctuations increases compared with
the previous estimates.

In the present paper we wish to critically reanalyze the results of
ref. \cite{jap}. We reconsider the role played by
the subcritical bubbles during the electroweak phase transition,
estimate their
size, amplitude and the formation rate.
Moreover, we study the influence of
subcritical bubbles on the formation of a
critical bubble by comparing the average subcritical energy
density associated with the fluctuation to
the energetics of barrier penetration in the thin wall approximation.
Our final conclusion, which differs from ref. \cite{jap}, is that
subcritical bubbles do not affect
nucleation of critical bubbles in an appreciable way.

The  paper is organized as follows. In Sect. {2} we briefly
describe the small supercooling limit and the thin wall app\-roxim\-ation.
In Sect. {3} we discuss the salient features of subcritical fluctuations and
present our Ansatz subcritical bubbles.
Thermal averages are performed in Sect. 4, and the crucial role of
thermalization is discussed in Sect. {5}. Finally, our numerical results and
conclusions are given in Sect. {6}.

\Section {thin wall approximation} {Small supercooling limit and thin wall
approximation}

First order phase transition and bubble dynamics in the Standard Model
have lately been studied in much detail, and it has become increasingly clear
\cite{clear} that for Higgs masses
considerably heavier than 60 GeV, the electroweak phase transition
 is only of weakly first order. For Higgs mass
$m_H > 100$ GeV the calculations, both perturbative and lattice ones, confront
technical problems\footnote{It is conceivable that for such large
Higgs masses the electroweak phase transition is close to a second order
and does not proceed by critical bubble formation.}.
Therefore, in the paper at hand we use a
phenomenological Higgs potential for the order parameter $\phi$ suitable
for a simple description of a first order phase transition:
\be{potential}
V(\phi ) = \frac 12 m^2(T)\phi^2 - \frac 13 \alpha T \phi^3 + \frac 14
\lambda\phi^4,
\ee
where we have not determined the parameters perturbatively but fit them, when
needed, according to a recent two-loop calculation of the gauge invariant
effective
potential \cite{lattice}. Most of the dynamical properties of the electroweak
phase transition associated with the potential \eq{potential},
such as the smallness of the latent heat, the bubble
nucleation rate and the size of critical bubbles, have been discussed
in \cite{kari}. For the purposes of the present paper it suffices to
recall only some of the results.

First we need the size of the critical bubble. Assuming that there is
only little supercooling, as seems to be the case for the electroweak
phase transition, the bounce action can be written as
\be{bounce}
S/T = {\alpha\over \lambda^{3/2}} {2^{9/2}\pi\over 3^5}
{\bar\lambda^{3/2}\over
(\bar\lambda - 1)^2}~,
\ee
where $\bar\lambda (T)=9\lambda m^2(T)/(2\alpha^2T^2)$.
The cosmological transition temperature is determined
from the relation that the Hubble rate equals the transition rate $\propto
e^{-S/T}$,
yielding $S/T_f \simeq {\rm ln} (M_{Pl}^4/T_f^4) \simeq 150$,
where $T_f$ is the transition temperature. Thus we obtain from \eq{bounce}
\be{lambdabar}
\bar \lambda(T_f) \simeq 1 - 0.0442{\alpha^{1/2}\over\lambda^{3/4}}\equiv
1-\delta.
\ee
On the other hand, small supercooling implies that $1-\bar\lambda=\delta\ll 1$,
i.e. $\alpha \ll 500 \lambda^{3/2}$. Solving for $\bar\lambda$ in \eq{bounce}
yields
the transition temperature $T_f$. One finds
\be{mass}
m^2(T_f) = {2 \alpha^2\over 9\lambda}\:\bar\lambda(T_f) \: T_f^2~.
\ee
The extrema of the potential are given by
\be{extrema}
\phi_\pm (T) = {\alpha T\over 2\lambda}(1 \pm\sqrt{1 - 8\bar\lambda/9}).
\ee
Expanding the potential at the broken minimum $\phi_{+}(T)$ we find
\be{epsilon}
-\epsilon\equiv V(\phi_{+},T_f)={1\over 6}m^2(T_f)\phi_{+}^2-
{1\over 12}\lambda\phi_{+}^4=-0.00218\:{\alpha^{9/2}\over
\lambda^{15/4}}
T_f^4+{\cal O}\left(\delta^2\right).
\ee
The height of the barrier is situated at $\phi_{-}\simeq
\phi_{+}/2$ with $V\left(\phi_{-},T_c\right)\equiv V_{{\rm
max}}=\alpha^4 T_c^4/(144\:\lambda^3)$, where $T_c$ is the temperature
at which $V(0)=V(\phi_{+})$, given by the condition $m(T_c)^2=(2\:
\alpha^2\: T_c^2/9\: \lambda)$. As $T_c\simeq T_f$ we may conclude that
the thin wall approximation is valid if $-\epsilon/V_{{\rm max}}=
0.314\:\alpha^{1/2}/\lambda^{3/4}\ll1$, or
$\alpha\ll 10\lambda^{3/2}$. Thus  the small supercooling
limit is clearly satisfied if the thin wall approximation is valid.

To get the size of the critical bubble we still need the surface
tension. One easily finds
\be{surface}
\sigma=\int_{0}^{\infty}\:d\phi\sqrt{2\:V(T_c)}={2\:\sqrt{2}\:\alpha^3\over
91\:\lambda^{5/2}}\:T_c^3.
\ee
We define the critical bubble radius by extremizing the
bounce action. The result is
\be{Rc}
R_c = 13.4\, {\lambda^{3/4}\over\alpha^{1/2}m(T_f)}.
\ee
Therefore $R_c$ is much
larger than the correlation length $\xi(T_f) = 1/m(T_f)$ at the transition
temperature, as it should.

\Section {description of subc. bubbles} {Subcritical bubbles}

We begin this Section by  briefly describing the procedure
adopted in ref. \cite{jap}. The following spherical Ansatz was chosen
for the subcritical bubble configuration:
\be{japans}
\phi_g(r)=\phi_{+}\:{\exp}\left[-{r^2\over R(t)^2}\right],
\ee
where $R(t)$ is the time dependent size of the subcritical bubble and $r$ is
the radial coordinate. In
order to estimate the typical size of a subcritical bubble, one defines the
canonical momentum $p_R\equiv \partial L_{{\rm eff}}(\dot{R},R)/\partial
\dot{R}$, whence an effective Hamiltonian
\be{hamiltonian}
H_{{\rm eff}}(
R,p_R)\equiv p_R \dot{R}-L_{{\rm eff}}
\ee
can be constructed.

The thermal average of $R$ was defined as
\be{japR}
\langle R \rangle={\int dp_R\: dR\: R\:
e^{-\beta H_{{\rm eff}}}\over
\int dp_R\: dR\:
e^{-\beta H_{{\rm eff}}}},
\ee
which, in the neighborhood of the critical temperature,
turns out to be much smaller than $\xi(T_c)$ .

To find the thermal fluctuation amplitude on the scale $\langle R
\rangle$, the following trial function was adopted
\be{japtrial}
\phi(r)=a\:{\exp}\left[-{r^2\over\langle R
\rangle^2}\right],
\ee
whence the amplitude in the symmetric vacuum was given by
\be{japamplitude}
{\langle a^2 \rangle} ={\int da\:a^2
e^{-\beta F\left(a,T\right)}\over
\int da\:
e^{-\beta F\left(a,T\right)}}.
\ee
At $T\simeq T_c$ the amplitude of a fluctuation $\sqrt{\langle a^2 \rangle}$
exceeded the inflection point $\phi_{-}(T_c)$ and
from this fact it was concluded that thermal fluctuations
drastically change the dynamics of the phase transition from the
ordinary first order type with supercooling.

We now wish to make some critical comments on the approach
used in ref. \cite{jap}
and present a remedy to it, which, as we shall argue, is more
realistic.

Let us first make the general observation that
it is the actual transition temperature $T_f$ rather than the critical
temperature $T_c$
which is relevant for the study on subcritical bubbles.
This is true in the sense that if critical bubbles are not important
at $T_f$, they most certainly will not be so at $T_c$. As we shall show,
it actually turns out that subcritical bubbles are not important even
at  $T_f$. This justifies, in retrospect, our choice $T=T_f$ for performing
the calculations.

In the case of a
weak first order phase transition the critical bubble is typically well
described by a thin wall approximation, where the  configuration is by
no means gaussian, but has a flat 'highland' (with $\phi$ determined
by the non-zero minimum of the potential) and a steep slope down to
$\phi=0$. Therefore it seems natural that also a large subcritical
bubble should resemble the critical one, {\it i.e.} when $R$ increases,
the form of the subcritical bubble should deform smoothly so that, when
$R=R_c$, the bubble is a critical one. This behaviour is
not reproduced by the Ansatz of Shiromitzu \etal, given
in \eq{japans}. Moreover, it has
the strange property
that in the limit $R\rightarrow 0$, the bubble becomes infinitely sharp
because the amplitude of the fluctuation is kept fixed, while, in
general, one would expect that the amplitude of the fluctuations depends
sensitively on their spatial size. Finally, but no less important,
the calculation of $\sqrt{\langle a^2\rangle}$ , \eq{japamplitude},
is performed by substituting for the typical size of the subcritical bubble
the average
$\langle R\rangle$ as calculated from \eq{japR}, which {\it only} is  valid
for a particular value of $a$, namely $a=\phi_{+}$. In other words, in
ref. \cite{jap} the only statistical degree of freedom is taken to be
the size of subcritical bubbles.

Motivated by these observations, let us  define a
subcritical bubble as a functional of both the amplitude $a$
and the radius $R$. For this purpose one has first to
study the behaviour of the
potential as a function of the amplitude. At $T_f$ there is a interval $\phi
\in [a_-,\ a_+]$ where $V(\phi ) \le 0$. If the amplitude of the bubble is
in that interval, there exists a
critical bubble-solution of the bounce action. This means that
we have a relation
$R_c = R_c(a)$ which reproduces \eq{Rc} if $a = \phi_+$. Therefore
$R_c(a)$
serves as an upper limit for the bubble radius in that region.

Moreover, as argued above, the
subcritical bubble should  be of the thin wall type,  with the central
region having the field value somewhere in the interval $\phi
\in [a_-,\ a_+]$.

These considerations lead us to
define different Ans\" atze for various regions in the $(a, R)$ -plane. When
$\phi\in [a_-,\ a_+]$, we use an Ansatz such that when
$R \rightarrow R_c(a)$, the
field configuration goes towards the thin wall form. For small $R$ we use a
simple gaussian configuration. For other values of $a$ we always take a
thin-wall like Ansatz. Thus we write for $\phi \in [a_-,\ a_+]$ and $R \leq
R_c(a)$
\be{ansatz1}
\phi(t,\ R) = a(t) \left[{R_c - R\over R_c}\phi_g + {R\over
R_c}\phi_t\right],
\ee
where $t$ is the time coordinate\footnote{Note, however, that we need not to
specify the explicit time evolution of $a$ and $R$ when dealing with
statistical averages.} and
\bea{functions}
\phi_g(R) &=& e^{-r^2/R^2},\\
\phi_t(R) &=& 1/(e^{m(r - R)} + 1).
\eea
Such an Ansatz reproduces the requirement that when $R\rightarrow R_c$,
subcritical
bubbles should resemble  critical ones. In practise the statistical averages
depend only weakly on $a$ because the main contribution to them comes from
the region of small $a$ and large $R$. Therefore we assume
for simplicity that criticality depends only weakly on $a$ and take
$R_c(a)=R_c$ to be a constant whenever possible.

For $\phi \not\in [a_-,\ a_+]$ we assume that no gaussian component is
present and write simply
\be{ansatz2}
\phi(t,\ R) = a(t)\phi_t(R).
\ee
However, the statistical averages are excepted to be quite insensitive of
the precise form of the bubble.

These Ans\" atze can be plugged into the action
\be{action}
S[a,\ R] = \int d^4x\, [\frac 12 (\partial\phi)^2 - V(\phi)]
\ee
from which the Lagrangian in terms of the dynamical variables $a$ and $R$ can
be extracted. In the practical calculation we have, whenever possible,
approximated $\phi_t$ by the step function.
After that is a simple matter to calculate the effective
Hamiltonian function $H_{{\rm eff}}$ of the
dynamical variables $a$ and $R$.
\Section {statistical averages etc.} {Statistical averages}

Once we have the Hamiltonian, we may calculate the
statistical average of a dynamical variable of the type $F(a, R)$ simply by
\be{generalave}
\langle F(a, R)\rangle = {\int dp_R\, dp_a\, da\, dR \,F(a, R)
e^{-\beta H_{{\rm eff}}}
\over \int dp_R\, dp_a\, da\, dR\, e^{-\beta H_{{\rm eff}}}}.
\ee
However, because the effective Lagrangian is of the form
\be{lag}
{\cal L}_{{\rm eff}} = \frac 12 \pmatrix{\dot a & \dot R\cr}
K \pmatrix{\dot a \cr \dot R\cr} - {\cal V},
\ee
where $K=K(a,R)$ is a symmetric matrix, after the momentum integration the
average
can be cast into the form
\be{ave}
\langle F(a, R)\rangle = {\int da\, dR \,F(a, R)
\sqrt{\det K} e^{-\beta {\cal V}}\over \int da\, dR\, \sqrt{\det K} e^{-\beta
{\cal V}}}.
\ee
The matrix
\be{K}
K = 4\pi \pmatrix{K_{11}& K_{12}\cr
           K_{21}& K_{22}\cr}
\ee
and the pseudopotential ${\cal V}$ are given separately for the two
regions. For $\phi \in [a_-,\ a_+]$ we obtain
\bea{Kinreg.1}
R_c^2\, K_{11} &=& \Delta^2 R^3 A^2_2 + 2 \Delta R^4 B_2^1 + \frac 13 R^5
\nonumber\\
R_c^2 \,K_{12} &=&  2 a \Delta^2 R^2 A_4^2 - a \Delta R^3 A_2^2 + \frac 13 a
R^4 +
a R^5 m I(mR) \nonumber\\
& & + a \Delta R^3 B_2^1 - a R^4 B_2^1 + 2 a \Delta R^3 B_4^1
+a \Delta R^4 m J_2(mR)\nonumber \\
R_c^2\, K_{22} &=& 4 a^2 \Delta^2 R A_6^2 + a^2 R^5 m^2 I(2mR) + 4 a^2 \Delta
R^3 m
J_4(mR) \nonumber\\
& & + a^2 R^3 A_2^2 - 2 a^2 R^3 B_2^1
+ \frac 13 a^2 R^3 - a^2\Delta R^2A_4^2 \nonumber\\
& & - 2a^2 R^4mJ_2(mR) + 4a^2
\Delta R^2 B_4^1 + 2a^2 R^4 m I(mR)
\eea
and
\bea{Linreg.1}
{R_c(a)^2\over 4\pi}{\cal V} &=&  2 a^2\Delta^2 R A_4^2 + 2 a^2 \Delta R^3 m
J_3(mR) + \frac 12a^2 R^5 m^2 I(2mR)\nonumber\\
 & & + \frac 12 m^2 a^2 \Delta^2 R^3 A_2^2 +
m^2 a^2 \Delta R^4 B_2^1 + \frac 16 m^2a^2 R^5 -
\frac 13 \alpha T {a^3\over R_c(a)}\Delta^3R^3A_2^3\nonumber\\
& & - \alpha T {a^3\over R_c(a)}\Delta^2 R^4 B_2^2 - \alpha T {a^3\over R_c(a)}
\Delta R^5 B_2^1 - \frac 19 \alpha T {a^3\over R_c(a)}  R^6\nonumber\\
& & + \frac 14 \lambda {a^4\over R_c(a)^2} \Delta^4 R^3 A_2^4
 + \lambda {a^4\over R_c(a)^2} \Delta^3 R^4 B_2^3
+ \frac 32 \lambda {a^4\over R_c(a)^2} \Delta^2 R^5 B_2^2\nonumber \\
& & + \lambda {a^4\over R_c(a)^2} \Delta  R^6 B_2^1
+ \frac 1{12} \lambda {a^4\over R_c(a)^2}  R^7.
\eea
Note that in \eq{Linreg.1} the $a$ -dependence of $R_c$ has to be used
explicitly
because the critical behavour is determined from it.
For the region where $\phi \not\in [a_-,\ a_+]$ the corresponding functions
are given by
\bea{Kinreg.2}
K_{11} &=& \frac 13 R^3 \nonumber\\
K_{12} &=& a R^3 m I(mR)\nonumber \\
K_{22} &=&  a^2 R^3 m^2 I(2mR)
\eea
and
\be{Linreg.2}
{1\over 4\pi}{\cal V} =  \frac 12 a^2 R^3 m^2 I(2mR) + \frac 16 m^2a^2 R^3 -
 \frac 19 \alpha T a^3  R^3 + \frac 1{12} \lambda a^4  R^3.
\ee
A number of shorthand notations have been introduced in the previous equations:
\bea{notations}
\Delta &=& R_c(a) - R\\
A_n^k &=& \int_0^\infty du\, u^n e^{- k u^2} = {\Gamma({n+1\over 2})
\over 2 k^{n+1\over 2}}\\
B_n^k &=& \int_0^1 du\, u^n e^{- k u^2} \\
I(x) &=& \int_0^1 du\, u^2 e^{x(u -1)} = \frac 1x - \frac 2 {x^2} +
\frac 2{x^3} - \frac 2{x^3} e^{-x}\\
J_n(x) &=& \int_0^1 du\, u^n e^{- u^2 + x(u-1)}.
\eea

\Section {upper bound for $R$}{Thermalization}

Motivated by the fact that thermal fluctuations can generate configurations
with spatial size comparable to the critical bubble radius, which may
affect the dynamics of a first order phase transition,
the authors of ref.
\cite{ElmforsEV} have estimated the lifetime of fluctuations
of an on-shell
Higgs field with zero  momentum $\left(p_0=m(T),{\bf
p}=0\right)$. This choice  reflects the fact
that critical bubbles are typically
much larger than the interparticle distance $\simeq 1/T$ in plasma.
Writing $p_0\equiv \omega-i\:\gamma/2$, one finds that the dispersion
relation is
\be{dis}
\omega^2=\left|{\bf p}\right|^2+m^2(T)+{1\over 4}\gamma^2,
\ee
where
\be{gamma}
\gamma={{\rm Im}\:\Gamma^{(2)} \over\omega},
\ee
$\Gamma^{(2)}$ being the two-point function for the Higgs field.

The imaginary part arises at one loop level, but because of kinematical
constraints, the two loop contribution is actually dominant in the
region of physical couplings.
The thermalization rate $\gamma$ for small
amplitude scalar fluctuations and large spatial size, $
R\sim\left|{\bf p}\right|^{-1}\gg \gamma^{-1}$, is estimated \cite{ElmforsEV}
to be of the order $\gamma
\simeq 10^{-2}\: T$ near the critical temperature, {\it i.e.}
much larger than the typical first order transition time. This means that all
small amplitude fluctuations with size larger than
\be{maxR}
R_{{\rm max}}={\cal O}(1/\gamma)
\ee will effectively
be absent from the mixture of subcritical bubbles and must
not counted in the thermal averages. In
practise, the limit \eq{maxR} is of the order of few times $R_c$, depending on
the actual value of $\gamma $. Even if it is not precisely known, its
inclusion in the calculations is important. Without it all statistical
averages would be dominated by infinite, infinitesimally small fluctuations.
Technically this can be seen from the \eq{ave}, where the integrals
diverge in the limit $a\goto 0,\ R\goto\infty$. It is important to note that
the divergence is not a problem of our Ansatz but merely a more general
phenomenon, which seems to be related to the general infra-red instability
problems emerging in the calculations of the effective action.

\Section {Results}{Results and conclusions}
We have computed the average radius and the amplitude of fluctuations
at $T=T_f$ from \eq{ave}
numerically, using a cut-off $R_{{\rm max}}$ as discussed in the
previous Section. For definiteness, we fitted our phenomenological potential
\eq{potential} to the two loop result for the effective
potential calculated in \cite{lattice} for the Higgs mass $M_H=70$ GeV.
This yields $\alpha\simeq 0.048$ and $\lambda\simeq 0.061$.
One readily verifies that we are indeed safely in the thin wall limit.
With these parameter values, in units of $R_c$, $T_f=85.70$,
$m^2(T_f)=56.78$ and $\gamma\simeq 10^{-2}\:T_f\simeq 0.875$.

Choosing, for instance,
$R_{{\rm max}}\simeq 6.6\:R_c$, we find (in units of $R_c$)
\be{results}
\langle a\rangle=1.16;~~
\langle R\rangle =2.93;~~
\langle a^2\rangle =12.6;~~
\langle R^2\rangle =12.2~.
\ee
Note that $\xi=m^{-1}(T_f)\ll \langle R\rangle$, so that the average
subcritical bubbles are much larger than the correlation size, as
seems reasonable. Indeed, the correlation length gives the order
of magnitude of the size of a (thermal) quantum fluctuation, so that, since
to build up a bubble configuration many quanta are needed, one can
expect that $\langle R\rangle$ is larger than $\xi(T_f)$. Here we remind
the reader that the authors of ref. \cite{jap} find the opposite result,
which is somehow suspect when the above considerations are taken into
account.

As far as the production rate for subcritical bubbles is concerned, it
is roughly given by \cite{gleiser}
\be{rate}
\Gamma_{sc}\simeq T_f\: {\rm exp}\left(-S/T_f\right),
\ee
where $S$, given by \eq{action}, is the free energy for
the  subcritical configuration.

We have found  the average rate for subcritical bubble formation
to be much larger than the Hubble rate $H$:
\be{scrate}
\Gamma_{sc}= T_f\langle e^{-S/T}\rangle \simeq 0.49\:T_f\gg H~.
\ee
Thus we find out that the subcritical bubbling is so fast that
effectively the background, on which a critical bubble is to formed,
is filled by subcritical fluctuations. In principle, this could
affect the way by which critical bubbles are produced and naively, one
could think that the extra energy would facilitate barrier penetration
or possibly even invalidate our thin wall approach. That this is not
the case can be ascertained by considering the average fluctuation energy
about $\phi=0$. Using \eq{results} one obtains
\be{vertailu}
\langle V(a)\rangle\simeq \frac 12 m^2(T_f)\langle a^2\rangle =357.7
\ll -\epsilon \simeq 5\times 10^3~,
\ee
where \eq{epsilon} and the fact that $\phi_+\simeq
48.05$ have been used. Also the average amplitude is clearly smaller than
the inflection point, $\sqrt{\langle a^2\rangle} \ll \phi_{-}$. Thus, in
terms of
barrier penetration, subcritical fluctuations represent only a
minor correction and they have no remarkable effect on the thin wall
approximation.

We should point out that the results \eq{results}
are rather sensitive to $R_{{\rm max}}$; for instance, if
we took $R_{{\rm max}}\simeq 3.3\:R_c$, we would gotten $\langle a\rangle=
1.94,~~\langle R\rangle =1.53$, together with an almost twice as large $
\langle a^2\rangle$ and an $\langle R^2\rangle$ smaller by a factor of
about four. This is a general trend: as $R_{{\rm max}}$ increases,
$\langle a\rangle$ and $\langle a^2\rangle$ decrease so that the
amplitude becomes more peaked, while $\langle R\rangle$ and
$\langle R^2\rangle$ grow. To explain analytically, for instance, the
scaling of $\langle R\rangle$ with the changing of $R_{{\rm max}}$
one can consider
the limit $R_c\rightarrow \infty$ in \eq{ave}  which is
reasonable in the vicinity of $T_c$. In such a case, it is easy to show
that
\be{example}
\langle R\rangle \propto {\int_0^{R{{\rm max}}}dR\: R^{1/2}\over
\int_0^{R{{\rm max}}}dR\: R^{-1/2}}=R_{{\rm max}},
\ee
whereas a similar calculation proves that $\langle R^2\rangle\propto
R_{{\rm max}}^2$. It is then clear that why, increasing $R_{{\rm max}}$
by a factor two, $\langle R\rangle$ and $\langle R\rangle^2$ increase by
a factor two and four, respectively. Analogous considerations explain
the decreasing of $\langle a\rangle$ and $\langle a^2\rangle$ with
increasing $R_{{\rm max}}$.

Note, however, that, for reasonable values of the cut-off  $R_{{\rm max}}$,
the subcritical nucleation rate does not change,
nor the conclusion that $\langle V(a)\rangle\ll -\epsilon$.

To conclude, we have find that subcritical bubbles are not, at least
for Higgs masses less than about 100 GeV, important during the
onset of the electroweak phase transition. We have argued that
critical bubble formation proceeds essentially unchanged, and here
we disagree with the result of Shiromizu \etal \cite{jap}.
The main reason for this discrepancy is that in ref. \cite{jap}
the only statistical degree of freedom describing the properties of
subcritical bubbles is taken to be their size $R$, keeping
their amplitude fixed at the value $\phi_+$. This fact led the authors
of ref. \cite{jap} to conclude that $\langle R\rangle\ll
\xi(T_c)$ which appears not reasonable. Indeed, to build up a
classical configuration, many thermal quanta (whose typical size is
the correlation length) are needed and one would then expect that
$\langle R\rangle\gg
\xi(T_c)$. To recover such a behaviour, we have treated both the amplitude
and the critical size of subcritical bubbles as statistical degrees
of freedom. We have then  estimated their average size, amplitude and
formation rate taking into account the crucial
role played by thermalization, showing that
 subcritical bubbles do not affect the
nucleation of critical bubbles in an appreciable way. From this fact we
conclude that the electroweak baryogenesis scenarios associated
with a first order electroweak phase transition remain a viable
possibility to explain the primordial baryon asymmetry in the Universe.

\newpage

\end{document}